# Melting of chiral order in terbium manganate (TbMnO$_3$) observed with resonant x-ray Bragg diffraction

S W Lovesey[1,2], V Scagnoli[3,4], M Garganourakis[3], S M Koohpayeh[5], C Detlefs[6] and U Staub[3]

1. ISIS Facility, STFC Oxfordshire OX11 0QX, UK
2. Diamond Light Source Ltd, Oxfordshire OX11 0DE, UK
3. Swiss Light Source, Paul Scherrer Institut, CH 5232 Villigen PSI, CH
4. ETH Zürich, Institut für Quantenelektronik, W. Pauli Strasse 16, 8093 Zürich, CH
5. Institute for Quantum Matter, Department of Physics and Astronomy, Johns Hopkins University, Baltimore, Maryland 21218, USA
6. European Synchrotron Radiation Facility, BP 220, 38043 Grenoble Cedex 9, F

**Abstract** Resonant Bragg diffraction of soft, circularly polarized x-rays has been used to observe directly the temperature dependence of chiral-order melting in a motif of Mn ions in terbium manganate. The underlying mechanism uses the b-axis component of a cycloid, which vanishes outside the polar phase. Melting is witnessed by the first and second harmonics of a cycloid, and we explain why the observed temperature dependence is different in the two harmonics. Our direct observation of melting is supported by a solid foundation of evidence, derived from extensive studies of the azimuthal-angle dependence of intensities with both linear and circular polarization.

## 1. Introduction

An electronic state in which charge and magnetic polarizations coexist has been at the centre of materials science in the past decade. More work is needed to fully understand the phenomenon of multiferroicity, and to develop practical applications, notably controlling charges by applied magnetic fields and spins by applied voltages.

Two different mechanisms seem able to generate magnetically-induced polarization. First, exchange-striction appearing in nearly collinear spin structures [1, 2]. The coexistence of ferromagnetic and antiferromagnetic spin coupling introduces frustration in the system that is partially released by shifting the magnetic ions. Ions with antiparallel spins get closer, while ions having parallel spins are moved further apart. Such is the case of hexagonal HoMnO$_3$ [3] or orthorhombic RMn$_2$O$_5$ (R = Tb, Ho, Dy) [4, 5, 6]. Secondly, a magneto-electric effect can arise from the Dzyaloshinskii-Moriya interaction [7, 8]. The D-M interaction favours orthogonal spins, and it may induce canting of nominally parallel spins deprived of inversion symmetry. In a magnetically-induced multiferroic, such as TbMnO$_3$, the presence of a chiral magnetic structure induces polarization by shifting the oxygen atoms [9, 10].

The latter mechanism has generated debate in the community, specifically on the role of the ionic displacements. Katsura et al first argued that a novel mechanism, based on a spin super-current ≈ S$_1$ x S$_2$, was the source of the magneto-electric effect [11]. In this approach, the magneto-electric coupling is entirely arising from the non-collinear spin structure, without the need to invoke structural distortion that would break inversion symmetry (pure electronic contribution to electric polarization). Further theoretical work insisted on lattice distortions, and concluded that they should play a pivotal role in the physics of such materials [12, 13]. Experimentally the lattice distortions, if any, are minute and only recently with an elegant and sophisticated experiment Walker et al have shown that there are indeed lattice distortions in the polar phase of TbMnO$_3$, even if only of few femto-meters for the Tb ions [14]. Malashevich and Vanderbilt [15] argue that the pure electronic contribution suffers an accidental cancellation due to the rotation of the Mn-O octahedral, but it is a relevant player for other phases or other multiferroics. In light of the current debate, it is valuable to observe directly changes occurring in the electronic density at the onset of the multiferroic phase.

In the next section, we describe the experimental method, resonant x-ray Bragg diffraction. To unambiguously detect and study chiral properties of TbMnO$_3$ we exploit circular polarization in the primary beam of x-rays, and measure intensities of satellite reflections. In addition, our communication reports corresponding results observed with linear polarization. Figure 1 depicts states of polarization and the plane of scattering. Intensities have been gathered with changes to the orientation (azimuthal angle) and temperature of the sample.

The established magnetic structure for the multiferroic phase of TbMnO$_3$ is consistent with our data for the first harmonic, as shown in Section 3 [10, 14, 16]. We report equal amounts of data on the first and second harmonics of the Mn chiral order, which appear concomitantly with the magnetic order at 42 K and persists also in the polar phase appearing at 27 K. Thereafter, in Section 4, we develop a complete, atomic theory of diffraction by a circular cycloid, and prove its correctness for intensities at the first and second harmonics, by rigorous tests against our data collected as a function of the azimuthal angle in Section 5. Melting of chiral order is the main topic in Section 6. Conclusions are gathered in Section 7.

## 2. Sample preparation and experimental method

A TbMnO$_3$ single crystal was grown by the floating-zone technique using a four mirror image furnace. The starting materials for the preparation of feed rods for the floating zone crystal growth were Tb$_4$O$_7$ and MnO (99.99 % purity) obtained from Alfa Aesar. Stoichiometric amounts of raw materials were thoroughly ground together and then synthesized at 1200 C for 20 h in air with an intermediate grinding after 10 h. The powder was then compacted into a rod (typically 6 mm in diameter and 80 mm long), and sintered in a box furnace at 1450 C for 8 h in air. Large, stoichiometric and crack-free crystals were grown at 0.5 mm/h with rotation rate of 15 rpm for the growing crystal and 0 rpm for the feed rod under static argon. A sample was subsequently cut with facets parallel to the crystallographic axes and mounted on a sample holder with the b-axis perpendicular to the sample holder surface.

Experiments were performed with the RESOXS chamber [22] at the X11MA beamline [23] of the Swiss Light Source. Twin Apple undulators provide linear, horizontal, π, and vertical, σ, and circularly, right and left, polarized x-rays with a polarization rate close to 100 %, cf Figure 1. The sample was attached to the cold finger of an He flow cryostat with a base temperature of 10 K. Azimuthal scans were achieved by rotation of the single crystal, with an accuracy of approximately ± 5 degree.

Angular anisotropy in electronic structure can produce structurally forbidden Bragg reflections. Such is the case with dysprosium borocarbide, for example, and weak space-group forbidden reflections observed in Thomson scattering are reported by Adachi et al [Adachi H et al 2002 *Phys. Rev. Lett.* **89** 206401]. A resonant event in diffraction produces a helpful enhancement of intensity. In addition, resonant diffraction can rotate primary polarization, and thus depend on photon helicity, whereas Thomson scattering is diagonal in polarization states. Absence of translation symmetry also generates space-group forbidden reflections, or satellite reflections. Intensities we report in resonant Bragg diffraction by TbMnO$_3$ are consequences of departures from spherical symmetry in electronic structure and, also, the absence of translation symmetry in a chiral structure. In accompanying calculations of the scattering length symmetry considerations are placed at the forefront ahead of energy profiles, which are those for oscillators [].

A theory of resonant x-ray Bragg diffraction is laid out by Dmitrienko [Dmitrienko V E 1983 *Acta Crystallogr.* A**39** 29, 1984 *Acta Crystallogr.* A**40** 89], and Templeton and Templeton report the first relevant data, e.g., tetragonal $K_2PtCl_4$ (P4/mmm-type) and sodium bromate ($P2_13$-type) [24]. Formulations of resonant diffraction found in the cited papers use classical optics and physical properties of crystals with Cartesian tensors, as in the treatise by Nye [Nye J F *Physical properties of crystals* (Clarendon Press, Oxford, 1985)], and no attempt is made to calculate an energy profile. An atomic theory appeared shortly afterwards [17] in response to data for resonant Bragg diffraction by an incommensurate magnetic motif [a full report of experiments on magnetically ordered holmium is found in Gibbs D et al 1991 *Phys. Rev.* B**43** 5663]. Hannon et al [17] show that, electronic multipole events in x-ray diffraction provide sensitivity to magnetic properties of a material, an eye-opening revelation at the time of publication that is taken for granted today. For the sake of demonstration, Hannon et al tackle the formidable task of describing electronic structure at an atomic level of detail by imposing cylindrical symmetry, in which dipole and quadrupole contributions to scattering by a resonant ion are all generated from a single material-vector assigned to a magnetic dipole. In consequence, their x-ray scattering length is not universally applicable, unlike Dmitrienko's symmetry-based formulation of resonant Bragg diffraction by non-magnetic electronic structure []. Neither Dmitrienko [] or Hannon et al [] calculate energy profiles, which are mere factors in contributions to the scattering length labelled by electronic symmetry. Haverkort et al [30] estimate the factors in scattering lengths constructed with Cartesian tensors, in the footsteps of Dmitrienko [] and Hannon et al [], for various symmetries, using a multiplet crystal-field approach to electronic structure. Simulations by Cricchio et al [Cricchio F et al 2009 *Phys. Rev. Lett.* **103** 107202] of hidden order in $URu_2Si_2$ using density-function theory for itinerant electrons are analysed in terms of spherical tensors that are employed by us. Such simulations make possible an estimate of the absolute value of the scattering length, while we settle for ratios of contributions inferred from our data. Reviews of many applications of resonant x-ray Bragg diffraction, and advances in formulations, include [18, 19, 20].

## 3. Observations with linear polarization

Hannon et al [17] argued that electronic multipole events in x-ray diffraction provide sensitivity to magnetic properties of a material. To this end, they reduced electronic structure to a stick-model, by imposing cylindrical symmetry, in which dipole and quadrupole contributions to scattering are all generated from a single material-vector, **z**, assigned to a magnetic dipole. Expression (3) in [17] is the corresponding electric dipole-electric dipole (E1-E1) scattering length, namely,

$$f_{E1} = \mathbf{e'} \cdot \mathbf{e}\, F^{(0)} - i(\mathbf{e'} \times \mathbf{e}) \cdot \mathbf{z}\, F^{(1)} + (\mathbf{e'} \cdot \mathbf{z})(\mathbf{e} \cdot \mathbf{z})\, F^{(2)}. \qquad (1)$$

In (1), **e** (**e'**) describes the polarization state on the primary (secondary) x-rays, depicted in Figure 1, and $F^{(j)}$s have the dimension of length and contain resonance denominators. The scalar contribution to (1) is independent of the direction of the magnetic dipole, and it adds to simple charge scattering. The second, dipole contribution depends linearly on **z**, and it produces first-harmonic satellites in diffraction by a spiral antiferromagnetic motif. The quadrupole contribution in (1) depends quadratically on the magnetic dipole, and can produce second-harmonic satellites, and contributions to both the scalar part and Templeton and Templeton scattering at space-group forbidden reflections [24].

Using a magnetic structure-factor $(0, f_b, f_c)$ to describe a dipole moment in the b-c plane, [10, 25, 26] let us consider two different sample orientations. If the sample has the crystallographic a- and b-axis in the plane of diffraction ($\psi = 180°$, $\mathbf{z} = -(f_b, 0, f_c)$) we obtain from (1),

$$f_{\sigma'\sigma}(180°) = F^{(2)} f_c^2, \quad f_{\pi'\pi}(180°) = iF^{(1)} f_c \sin(2\theta) + F^{(2)} f_b^2 \cos^2\theta,$$

$$f_{\pi'\sigma}(180°) = -iF^{(1)} f_b \sin\theta + F^{(2)} f_b f_c \cos\theta. \qquad (2)$$

Here, $\theta$ is the Bragg angle shown in Figure 1. If the sample has the crystallographic b- and c-axis in the plane of diffraction ($\psi = 90°$, $\mathbf{z} = -(f_b, f_c, 0)$),

$$f_{\sigma'\sigma}(90°) = 0, \quad f_{\pi'\pi}(90°) = F^{(2)}(f_b^2 \cos^2\theta - f_c^2 \sin^2\theta),$$

$$f_{\pi'\sigma}(90°) = -iF^{(1)}(f_c \cos\theta + f_b \sin\theta). \qquad (3)$$

Terms in (2) and (3) linear in $f_b$, $f_c$ and quadratic in $f_b$, $f_c$ contributes to first-order and second-order satellites, respectively. These rules are not valid in the general case, because they omit T & T scattering in a first-order satellite.

The quantity $|f_{\pi'\pi}|^2$ is proportional to intensity in the $\pi'\pi$ channel and our data are shown in Figure 2, together with intensities in the remaining three channels. The selection rules established above hold for the first harmonic (0, $\tau$, 0), and confirm that the established magnetic structure for the multiferroic phase [10, 25, 27] is consistent our data. In particular, there is no intensity at $\psi = 180°$ in the $\sigma'\sigma$ channel and no intensity in the $\sigma'\sigma$ and $\pi'\pi$ channels at $\psi = 90°$. Different is the case for the second harmonic (0, $2\tau$, 0). At $\psi = 180°$ we have intensity in all four channels as expected, but at $\psi = 90°$ there is intensity in $\sigma'\sigma$, in contrast with prediction from expressions in (3). Having more intensity in the $\pi'\pi$ than $\sigma'\sigma$ discard lattice distortions as a possible source of the observed intensity. In order to explain such a surprising result we move to a full description of resonant diffraction by a circular cycloid, leaving behind the restriction to cylindrical symmetry and the stick-model.

### 4. Circular cycloid

The stick-model (1) provides but a useful glimpse at the electronic content of diffracted intensities, although it is often used without questioning its strengths and weaknesses. Five years after Hannon et al [17] communicated their ground-breaking insight they published a brief account of a complete, atomic theory of resonant x-ray diffraction [28], since when there have been discussions in complementary hues [19, 29, 30, 31]. Electronic degrees of freedom are encapsulated in irreducible, spherical multipoles, which cannot be represented by a material-vector, in the general case [29]. Our notation for a spherical multipole is $\langle T^K_Q \rangle$, with a complex conjugate $\langle T^K_Q \rangle^* = (-1)^Q \langle T^K_{-Q} \rangle$, where the positive integer K is the rank and Q the projection, which satisfies $-K \leq Q \leq K$ [19, 31]. Angular brackets $\langle ... \rangle$ denote the time-average of the enclosed quantum-mechanical operator, i.e., a multipole is a property of the electronic ground-state. Multipoles are parity-even for an E1-E1 event under consideration, and $(-1)^K$ their time signature.

Thermodynamic properties of multipoles serve to contrast the stick-model of the previous section and spherical multipoles. One can make the identification $\mathbf{z} = (x, y, z) \propto \langle T^1 \rangle$, and the dipole $\langle T^1 \rangle$ is known to be a linear combination of spin, $\langle S \rangle$, orbital moment, $\langle L \rangle$, and a dipole that expresses magnetic anisotropy [32]. As for the thermodynamic properties of $\langle T^1 \rangle$ it might reasonably scale with the total angular momentum, $\langle \mathbf{J} \rangle$. In the same vein, a quadrupole $\langle \mathbf{T}^2_0 \rangle$ might scale with $[3\langle (J_z)^2 \rangle - J(J + 1)]$, which vanishes in the absence of magnetic correlations and, also, when $J = 1/2$ (a decomposition of $\langle \mathbf{T}^2 \rangle$ into standard operators is available [33]). In the stick-model the corresponding quantity for the quadrupole is $z^2 \propto \langle J_z \rangle^2$. To illustrate likely differences in temperature dependences of quadrupole entities, Figure 3 displays normalized values of $\langle J_z \rangle^2$ and $[3\langle (J_z)^2 \rangle - J(J + 1)]$ as a function of temperature and J. Results are derived from an isotropic Heisenberg model treated with the molecular-field approximation [21].

Diffraction by an ideal circular-cycloid with moment rotation in the y-z plane is discussed by Scagnoli and Lovesey [34]. The super-cell length $L = (2n + 1)a$ where $a$ is a lattice spacing and $n$ an integer. The integer $f$ measures the wavevector in units of the fraction $(2\pi/a)(a/L) = 2\pi/(L(2n + 1))$, while the turn angle $= 2\pi/(2n + 1)$. Multipoles for the super-cell are denoted by $\langle C^K_Q \rangle$ and some expressions for $K = 1$ and $K = 2$ are listed in Table I. A few properties of $\langle C^K \rangle$ are more or less obvious, e.g., $\langle C^1 \rangle = 0$ for the second harmonic, $f = 2$, agrees with the stick-model (1), and a 90° phase shift between y- and z-dipoles in the ideal y-z cycloid, $\langle C^1_z \rangle = - i\langle C^1_y \rangle$. General results, and specific results for the ideal y-z cycloid, include: (i) Non-zero multipoles obey the identity for rotation by 180° about the axis normal to the plane of the cycloid, $C_{2x} \langle C^K_Q \rangle = (- 1)^K \langle C^K_{-Q} \rangle$. (ii) Using the identity $C_{2x} \langle C^K_Q \rangle = (- 1)^f \langle C^K_Q \rangle$ one finds $\langle C^K_{-Q} \rangle = (- 1)^f (- 1)^K \langle C^K_Q \rangle$. (iii) $\langle C^1_x \rangle \propto \langle C^1_{+1} - C^1_{-1} \rangle = 0$ because $\langle C^1_Q \rangle = 0$ for $f > 1$ and $\langle C^1_{+1} \rangle = \langle C^1_{-1} \rangle$ for $f = 1$. In the general case (iv) for given $f$ and $K$ all $\langle C^K_Q \rangle$ are proportional to one another. Scaling coefficients are complex and depend on both the magnitude and sign of the projection Q. (v) $\langle C^K_Q \rangle$ does not depend on n. (vi) $\langle C^K_Q \rangle$ is not Hermitian. (vii) $\langle C^K_Q \rangle = 0$ for $K < f$.

For the purpose of calculating unit-cell structure factors, F, it is always convenient to construct linear combinations of multipoles that are even ($A_{K,Q}$) and odd ($B_{K,Q}$) functions of the projection, Q [34] and,

$A_{K,Q} = A_{K,-Q} = [\exp(-iQ\alpha)_{K,Q} + \exp(iQ\alpha)_{K,-Q}]/2$,

$B_{K,Q} = - B_{K,-Q} = [\exp(-iQ\alpha)_{K,Q} - \exp(iQ\alpha)_{K,-Q}]/2$, (4)

with $\alpha = 90°$ for a Bragg wavevector parallel to the crystal b-axis. For a commensurate motif of multipoles, elements of symmetry in the space group will dictate the make-up of the electronic structure factor $_{K,Q}$ in (4).

To construct a minimal model of a cycloid in TbMnO$_3$ we seek guidance from the space group Pbnm in which Mn ions occupy sites 4b (standard setting Pnma, #62 [34]) and find,

$$\Psi_{K,Q} = \{1 + C_{2z} + \exp(if\beta)[C_{2x} + C_{2y}]\} \langle C^K_Q \rangle.$$

Here, $\beta$ is the fractional wave-vector and $C_{2\eta}$ denotes the operation of rotation by 180° about the $\eta$-axis. To make use of this expression we have to determine quantities $C_{2z}\langle C^K_Q \rangle$ and $C_{2y}\langle C^K_Q \rangle$ for K = 1 and 2, while $C_{2x}\langle C^K_Q \rangle = (-1)^f \langle C^K_Q \rangle$ for the y-z cycloid. Using specific results in Table I, we find $\Psi_{K,Q}$ proportional to (1 − exp(i$\beta$)) for f = 1, where $\beta = \pi\tau \approx 51°$. Because the spatial phase factor is common to all multipoles it cancels out in ratios, which all that can be inferred from diffraction data. With f = 2, operations $C_{2z}\langle C^2_Q \rangle$ and $C_{2y}\langle C^2_Q \rangle$ have the effect of creating the complex conjugate of $\langle C^2_Q \rangle$. In consequence, $\Psi_{2,Q} = 2(1 + \exp(2i\beta))$ Re.$\langle C^2_Q \rangle$ leading to the conclusion that real parts of multipoles $\langle C^2_Q \rangle$ contribute to scattering with f = 2.

In light of these findings we equate $\Psi_{K,Q}$ and $\langle C^K_Q \rangle$ in (4), and the expressions,

$$A_{K,Q} = \langle C^K_Q \rangle [\exp(-iQ\alpha) + \exp(iQ\alpha)(-1)^{K+f}]/2,$$
$$B_{K,Q} = \langle C^K_Q \rangle [\exp(-iQ\alpha) - \exp(iQ\alpha)(-1)^{K+f}]/2, \quad (5)$$

define our minimal model of a circular cycloid for Mn ions in terbium manganate. - (0, $\tau$, 0) and f = 1; allowed contributions are,

$$A_{1,0} = \langle C^1_0 \rangle, \; A_{2,1} = -i\langle C^2_{-1} \rangle, \; B_{1,1} = -i\langle C^1_{-1} \rangle \text{ and } B_{2,2} = -\langle C^2_{-2} \rangle. \quad (6)$$

The dipole $\langle C^1_0 \rangle \equiv \langle C^1_z \rangle = (1/2)[\langle T^1_0 \rangle - i\langle T^1_y \rangle]$ and $\langle C^2_{-1} \rangle = \langle C^2_{-2} \rangle$ can be complex. Previously, we noted the identities $\langle C^1_x \rangle = 0$ and $\langle C^1_z \rangle = -i\langle C^1_y \rangle$ for the y-z cycloid. We use $A_{1,0}$ to normalize multipoles inferred from data, Table II. The quantity t = $iB_{1,1}/A_{1,0}$ is proportional to $\langle C^1_y \rangle/\langle C^1_z \rangle$ and it is purely real for the ideal cycloid. We anticipate that quadrupole contributions are small. For $\langle C^2_{-1} \rangle$ is composed of charge fluctuations normal to the y-z plane, which are absent in the stick-model. Specifically, $\langle C^2_{-1} \rangle = (i\langle xy \rangle - \langle xz \rangle)/\sqrt{6}$, where $\langle \alpha\beta \rangle$ is a standard, purely real Cartesian quadrupole that is represented by $\langle \alpha \rangle \langle \beta \rangle$ in the stick-model.

Unit-cell structure factors for an E1-E1 event have the property that $F_{\sigma'\sigma}$ and $F_{\pi'\pi}$ do not depend on $B_{K,Q}$. We find (f = 1),

$$F_{\sigma'\sigma} = -i\sin(2\psi)A_{2,1}, \; F_{\pi'\pi} = (i/\sqrt{2})\sin 2\theta \cos(\psi)A_{1,0} - i\sin^2\theta \sin(2\psi)A_{2,1},$$
$$F_{\pi'\sigma} = -(i/\sqrt{2})\cos\theta \sin(\psi) A_{1,0} + i\sin\theta B_{1,1}$$
$$- i\sin\theta \cos(2\psi) A_{2,1} + i\cos\theta \sin(\psi) B_{2,2}. \quad (7)$$

A result for $F_{\sigma'\pi}$ is derived from $F_{\pi'\sigma}$ by the change of sign to both $A_{K,Q}$. Quadrupoles in (7) contribute Templeton and Templeton scattering by a cycloid

[24]. These contributions are omitted by Jang et al [10] without supporting evidence.

- (0, 2τ, 0) and f = 2; Recall that $\langle C^1 \rangle$ = 0, and allowed contributions are,

$$A_{2,0} = \langle C^2_0 \rangle, \quad A_{2,2} = -\langle C^2_{-2} \rangle, \text{ and } B_{2,1} = -i\langle C^2_{-1} \rangle. \tag{8}$$

The quadrupole $\langle C^2_0 \rangle = (3\langle zz \rangle + \langle x^2 - y^2 \rangle - 4i\langle yz \rangle)/8$ is complex, while ratios of multipoles are purely real, with $\langle C^2_{-1} \rangle/\langle C^2_0 \rangle = \sqrt{(2/3)}$ and $\langle C^2_{-2} \rangle/\langle C^2_0 \rangle = \sqrt{(1/6)}$. If we make $\langle C^2_0 \rangle$ a common factor, $F_{\sigma'\sigma}$ and $F_{\pi'\pi}$ are expected to be purely real. On the other hand, $F_{\pi'\sigma}$ contains an imaginary component due solely to $B_{2,1} = -i\langle C^2_{-1} \rangle$, and the sign of $B_{2,1}$ is not available from the intensity, $|F_{\pi'\sigma}|^2$. For the second harmonic (f = 2),

$$F_{\sigma'\sigma} = (1/2) \cos(2\psi)[\sqrt{(3/2)}A_{2,0} + A_{2,2}] - (1/2) [A_{2,2} - \sqrt{(1/6)}A_{2,0}],$$
$$F_{\pi'\pi} = (1/2) \sin^2\theta \cos(2\psi)[\sqrt{(3/2)}A_{2,0} + A_{2,2}]$$
$$\qquad + (1/2) (1 + \cos^2\theta) [A_{2,2} - \sqrt{(1/6)}A_{2,0}], \tag{9}$$
$$F_{\pi'\sigma} = -(1/2) \sin\theta \sin(2\psi) [\sqrt{(3/2)}A_{2,0} + A_{2,2}] - \cos\theta \cos(\psi) B_{2,1},$$

and $F_{\sigma'\pi}$ is derived from $F_{\pi'\sigma}$ by the change of sign to both $A_{2,Q}$. In contrast to (3), at the second harmonic $F_{\sigma'\sigma}$ can be different from zero at $\psi = 90°$. This finding is in accord with data in Figure 2.

## 5. Azimuthal-angle scans

- (0, τ, 0) Data are displayed in Figure 4. They are compared to predictions for the f = 1 circular cycloid with structure factors (7). Expressions for intensities in terms of structure factors, including circular polarization in the primary beam, can be found in reference [35]. Equation (10) is a specific example.

One can show that, $_{1,0}$ and $_{2,1}$ are proportional to $\langle T^1_0 \rangle$, which is purely real, and $\langle T^2_{+2} - T^2_{-2} \rangle \propto i\langle xy \rangle$, respectively. We have argued the case for the quadrupole $\langle xy \rangle$ to be small, which means $A_{2,1} \approx B_{2,2} \approx 0$ and $F_{\sigma'\sigma} \approx 0$. The ideal circular-cycloid yields $t = iB_{1,1}/A_{1,0} = 0.71$. Table II contains a value of $B_{1,1}/A_{1,0}$ inferred from a fit to data collected at 10 K, by when it has softened and amounts to 63% of its ideal value. Note that this ratio might depend slightly on energy as the different terms could have different energy dependences, in particular at the Mn $L_{2,3}$ edges. This will also lead to a possible azimuthal-angle dependent spectral shape.

- (0, 2τ, 0) Data are displayed in Figure 5, together with fits to intensities derived with expressions (9). Inferred values of $A_{2,2}/A_{2,0}$ and $B_{2,1}/A_{2,0}$ in Table II accord with predictions based on a circular cycloid, given in the previous section, with the magnitude of $A_{2,2}$ small compared to $|B_{2,1}|$ and a phase difference of 90° between the two contributions. In light of the conclusion drawn in the following section, reduction at 10K of $B_{2,1}/A_{2,0}$ to 87% of its magnitude in the ideal cycloid is consistent with (i) a partially melted cycloid and (ii) a dependence of intensity on temperature that is different for first and second harmonics. Figure 6 illustrates further the limitation of the stick-model. The azimuthal angle dependence of the (0, 2τ, 0) reflection is strongly dependent on the energy of the incident x rays. Terms in (2) and (3) quadratic in $f_b$, $f_c$ are not compatible with such complex dependence.

### 6. Observations with circular polarization

- (0, τ, 0) Dependence of the diffracted intensity as a function of circular polarization, $P_2$, is proportional to Im.$[(F_{\sigma'\pi})^*F_{\sigma'\sigma} + (F_{\pi'\pi})^*F_{\pi'\sigma}]$; see, e.g., [35]. Using expressions in (7) for the four unit-cell structure factors, the total intensity in units of $|A_{1,0}|^2$ is,

$$I = \sin^2\theta \, [\cos^2\theta + t^2 + \sqrt{2}\, t\, P_2 \cos\theta], \qquad (10)$$

for $\psi = 180°$. According to data in Figure 7, $t = -i\langle C^1_y\rangle/(\sqrt{2}\langle C^1_z\rangle)$ vanishes outside the polar phase where intensity no longer depends on $P_2$. This behaviour amounts to melting of Mn chiral order driven by softening of the dipole component of the cycloid that is parallel to the b-axis.

At the first harmonic, the temperature dependence of intensity is produced solely by dipoles provided corresponding quadrupoles are zero, or very small, and this appears to be the case from our data. The quadrupoles in question are entirely absent in a stick-model, used for the purpose of a first, simple demonstration by Hannon et al [17].

- (0, 2τ, 0) We anticipate $A_{2,2}/A_{2,0}$ purely real and $B_{2,1}/A_{2,0}$ purely imaginary. Data in Figure 7 for $\psi = 180°$ are consistent with these expectations. For dependence of the diffracted intensity on $P_2$ is controlled by Im.$(B_{2,1}/A_{2,0})$, according to the expressions (9), and the experimental evidence in Figure 7 is that it vanishes outside the polar phase. In this case, with $f = 2$, the dependence of intensity on temperature is produced by a ratio of quadrupoles. This fact accounts for a temperature dependence of intensity that is different to the first harmonic, and results displayed in Figure 3 for an isotropic Heisenberg magnet are indicative of such a difference.

## 7. Conclusions

Chiral order in a material is unambiguously detected by a probe with a matching characteristic, and we have used circularly polarized x-rays to detect such order in a single crystal of terbium manganate ($TbMnO_3$). Likewise, tuning the energy of x-rays to an atomic resonance of a manganese ion, the Mn $L_{2,3}$-edges, means these ions and no others participate in the chiral order. Diffraction as a function of rotation about the Bragg wavevector is entirely consistent with the diffraction pattern of a cycloid, using either primary linear or circular polarization. On this basis, we are able to demonstrate that melting of chiral order in the Mn motif, with increasing temperature, proceeds by a softening of the magnetic dipole-moment parallel to the b-axis of the crystal, and chiral order is absent outside the polar phase. Melting is observed in our data at both the first and second harmonics of the cycloid. Ab initio calculations of the energy dependence of the second harmonic could shed more light on the modification of the electronic structure occurring at the onset of the polar phase.

On the way, we make good shortcomings of a model of resonant x-ray diffraction originally exploited by Hannon et al [17] for its extreme simplicity, but that can lead to misleading results if applied with little or no questioning.

**Acknowledgement**

We are grateful to Professor E Balcar for his preparation of Figures 1 and 3. We have benefited from the experimental support of the X11MA beam-line staff. The financial support of the Swiss National Science Foundation and its NCCR MaNEP is gratefully acknowledged. Work at IQM was supported by DoE, Office of Basic Energy Sciences, Division of Materials Sciences and Engineering under Award DE-FG02-08ER46544.

**References**
1] Cheong S W and Mostovoy M 2007 *Nature Mat.* **6** 19
2] Picozzi S et al 2007 *Phys. Rev. Lett.* **99** 227201
3] Vajk O P et al 2005 *Phys. Rev. Lett.* **94** 087601
4] Hur N et al 2004 *Phys. Rev. Lett.* **93** 107207
5] Chapon L C et al 2004 *Phys. Rev. Lett.* **93** 177402
6] Blake G R et al 2005 *Phys. Rev. B* **71** 214402
7] Dzyaloshinsky I 1958 *J. Phys. Chem. Solids* **4** 241
8] Moriya T 1960 *Phys. Rev. Lett.* **4** 228

9] Sergienko I A and Dagotto E 2006 *Phys. Rev. B* **73** 094434
10] Jang H et al 2011 *Phys. Rev. Lett.* **106** 047203
11] Katsura H et al 2005 *Phys. Rev. Lett.* **95** 057205
12] Malashevich A and Vanderbilt D 2008 *Phys. Rev. Lett.* **101** 037210
13] Xiang H J et al 2008 *Phys. Rev. Lett.* **101** 037209
14] Walker H C et al 2011 *Science* **333** 1273
15] Malashevich A and Vanderbilt D 2009 *Phys. Rev. B* **80** 224407
16] Mannix D et al 2007 *Phys. Rev. B* **76** 184420
17] Hannon J P et al 1988 *Phys. Rev. Lett.* **61** 1245; *ibid* 1989 **62** 2644 (E)
18] Dmitrienko V E et al 2005 *Acta Crystallogr.* A**61** 481
19] Lovesey S W et al 2005 *Phys. Rep.* **411** 233
20] Matsumura T et al 2013 *J. Phys. Soc. Japan* **82** 021007
21] Lovesey S W 1993 *Rep. Prog. Phys.* **56** 257
22] Staub U 2008 *J. Synchr. Rad.* **15** 469
23] Flechsig U 2010 *The 10th international conference on synchrotron radiation instrumentation* AIP Conference Proceedings, **1234** 319 (AIP, 2010)
24] Templeton D H and Templeton L K 1985 *Acta Crystallogr.* A**41** 133; 1985 *Acta Crystallogr.* **41** 365; *ibid* 1986 **42** 478
25] Kenzelmann M et al 2005 *Phys. Rev. Lett.* **95** 087206,
26] Brinks H W et al 2001 *Phys. Rev. B* **63** 094411
27] Wilkins S B et al 2009 *Phys. Rev. Lett.* **103** 207602
28] Luo J et al 1993 *Phys. Rev. Lett.* **71** 287
29] Collins S P and Bombardi A *Magnetism and Synchrotron Radiation* (Springer, Heidelberg, 2010) Springer Proceedings in Physics, Vol. 133
30] Haverkort M W et al 2010 *Phys. Rev. B* **82** 094403
31] Lovesey S W and Balcar E 2013 *J. Phys. Soc. Japan* **82** 0210082
32] Carra P et al 1993 *Phys. Rev. Lett.* **70** 694
33] Lovesey S W and Balcar E 1997 *J. Phys.: Condens. Matter* **9** 8679
34] Scagnoi V and Lovesey S W 2009 *Phys. Rev. B* **79** 035111
35] Lovesey S W et al 2011 *Phys. Rev. B* **83** 054427

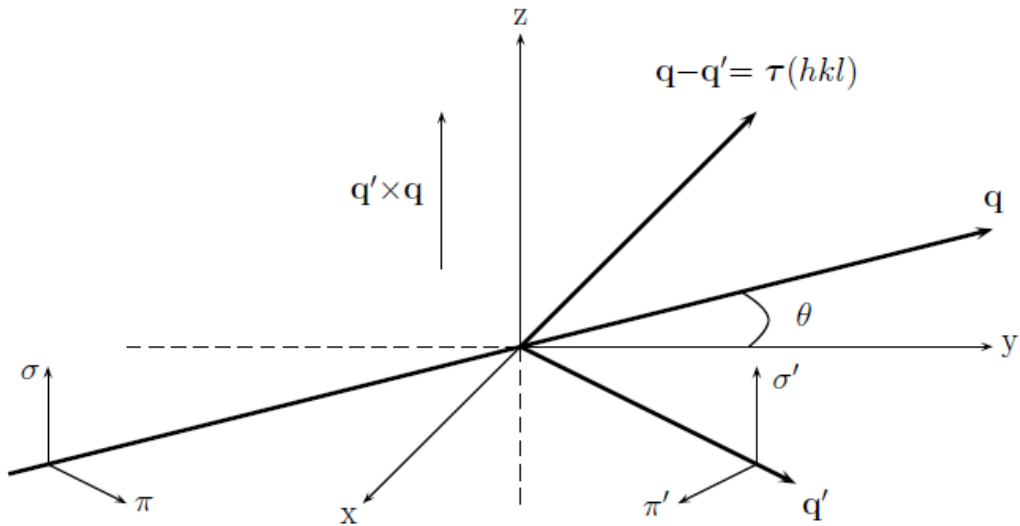

Figure 1. Cartesian coordinates (x, y, z) and x-ray polarization and wavevectors. The plane of scattering spanned by primary (**q**) and secondary (**q'**) wavevectors coincides with the x-y plane, and the primary beam is deflected through an angle 2θ. Polarization labelled σ and σ' is normal to the plane and parallel to the z-axis, and polarization labelled π and π' lies in the plane of scattering. Crystal axes a, b, c coincide with x, y, z in the nominal setting of the sample (TbMnO$_3$).

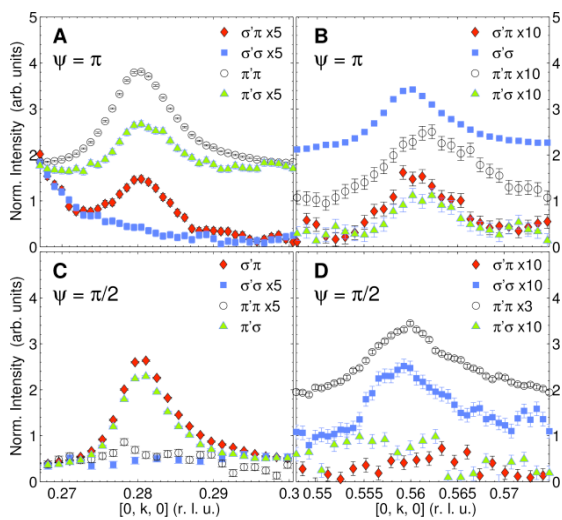

Figure 2. (Colour online) Polarization analysis at the Mn $L_2$-edge. Open (black) circle represents intensities collected in the $\pi'\pi$ channel, closed (blue) square in the $\sigma'\sigma$ channel. Closed (red) diamond and closed (green) triangle represent intensities in the rotated channels, $\sigma'\pi$ and $\pi'\sigma$, respectively. Data were collected at two different azimuthal angles and at two satellite reflections. A) $(0, \tau, 0)$ with $\psi = 180°$. B) $(0, 2\tau, 0)$ with $\psi = 180°$. C) $(0, \tau, 0)$ with $\psi = 90°$. D) $(0, 2\tau, 0)$ with $\psi = 90°$. Data are shifted for clarity.

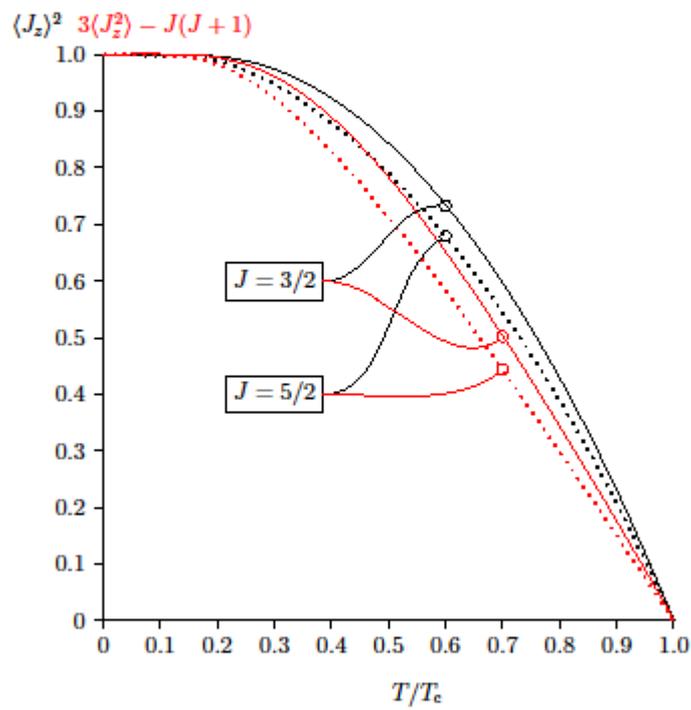

Figure 3. Values of the normalized quantities $\langle J_z\rangle^2/J^2$ (shown with black lines) and $[3\langle (J_z)^2\rangle - J(J + 1)]/[J(2J - 1)]$ (shown in red) as a function of reduced

temperature, $T/T_c$, and J. Results are derived from an isotropic Heisenberg model treated with the molecular-field approximation. Expressions for these quantities are found in reference [21].

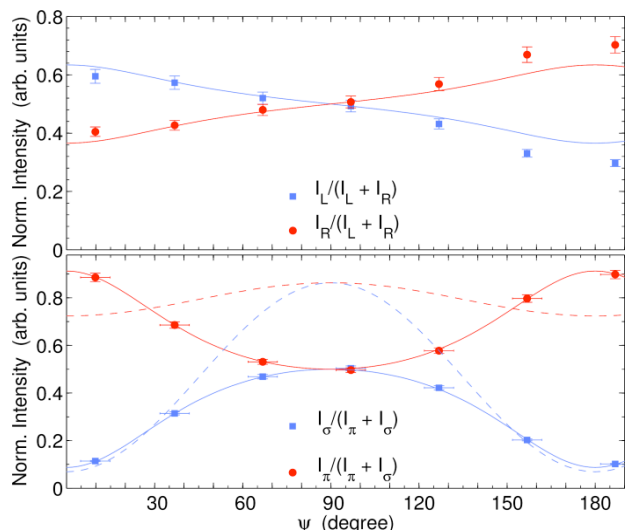

Figure 4. Azimuthal-angle dependence of the $(0, \tau, 0)$ reflection at 10 K. The energy of the primary x-rays (653 eV) corresponds to the Mn $L_2$-edge. Top panel: Dependence measured as a function of the circular polarization (helicity) of the incident x-rays: normalized intensities $I_L/(I_L + I_R)$ and $I_R/(I_L + I_R)$. Lower panel: Dependence measured with linearly polarized incident x-rays: normalized intensities $I_\pi/(I_\pi + I_\sigma)$ and $I_\sigma/(I_\pi + I_\sigma)$. Continuous lines are fits to intensities derived from (7). For completeness, dashed lines represent the intensities $I_\pi$ and $I_\sigma$ calculated from (7).

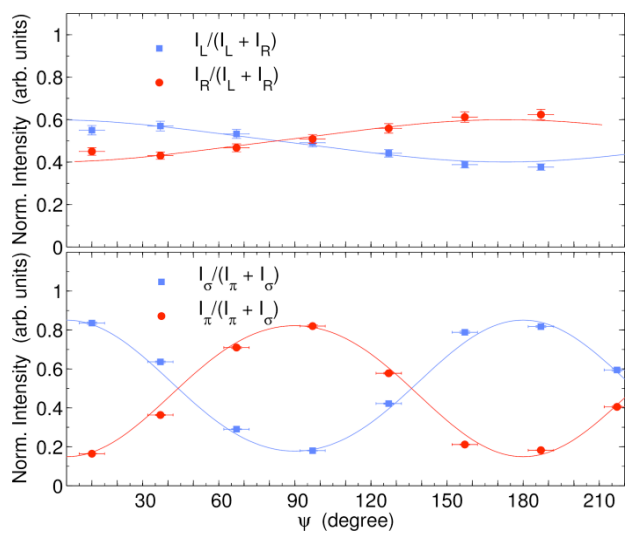

Figure 5. Azimuthal-angle dependence of the (0, 2τ, 0) reflection at 10 K. The energy of the incident x-rays (653 eV) corresponds to the Mn L$_2$-edge. Top panel: Dependence measured as a function of the helicity of the incident x-rays: I$_L$, I$_R$, normalized by their sum, as in Figure 4. Lower panel: Dependence measured with linearly polarized incident x-rays: I$_\pi$ and I$_\sigma$ normalized by their sum, as in Figure 4. Continuous lines are fits to intensities derived from (9).

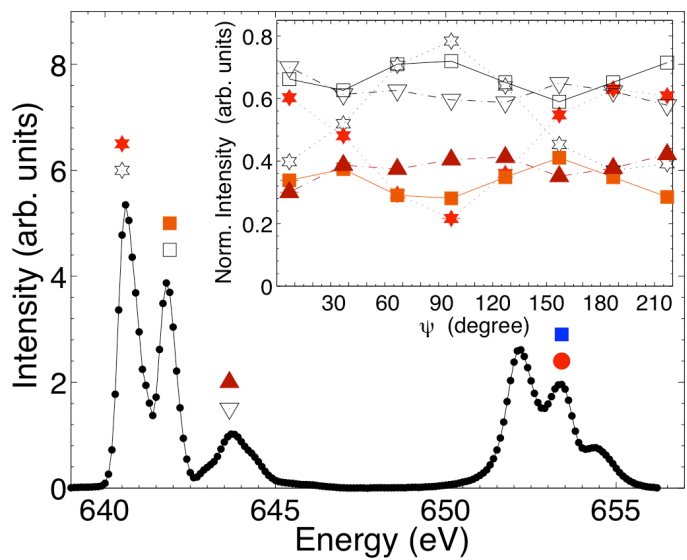

Figure 6. (Colour online) Energy dependence of the (0, 2τ, 0) reflection with π incident x rays. Inset: azimuthal angle dependence of the (0, 2τ, 0) reflection at 10 K at different energies. The energy of the incident x-rays corresponds to the Mn L$_3$ edge at 640.5 eV (Six-pointed star), 641.9 eV (square) and 643.65 eV (triangle), respectively. Dependence measured with linearly polarized incident x-rays: I$_\pi$/(I$_\pi$ + I$_\sigma$) in red and I$_\sigma$/(I$_\pi$ + I$_\sigma$) in black. Lines are guide to the eye.

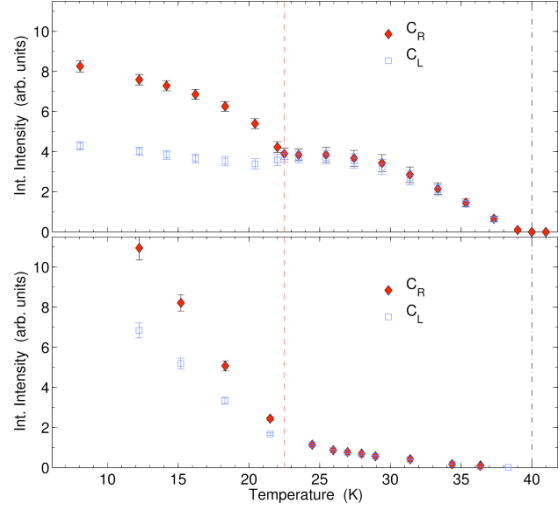

Figure 7. (Colour online) Top panel: Temperature dependence of the (0, τ, 0) reflection. Lower panel: Temperature dependence of the (0, 2τ, 0) reflection. Open (blue) square are integrated intensities collected with left-circular polarization $C_L$, and Stokes parameter $P_2 < 0$. Closed (red) diamond represent integrated intensities collected with right-circular polarization $C_R$, and Stokes parameter $P_2 > 0$. The temperature dependence was collected at the azimuthal angle $\psi = 180°$ which corresponds to the a-axis in the scattering plane. The energy of the incident x-rays was 653 eV.

**Table I**. Multipoles with K = 1 and 2 for the first (f = 1) and second (f = 2) harmonic of a circular cycloid with moment rotation in the y-z plane. Results are generated from an expression given by Scagnoli and Lovesey [34]. (Note that in equation (F4) the sign in front of the second double sum is incorrect and it should be a minus sign).

f = 1.  $\langle C^1_0 \rangle = (1/2)[\langle T^1_0 \rangle - i\langle T^1_y \rangle]$ with $\langle T^1_y \rangle = i\langle T^1_{-1} + T^1_{+1} \rangle/\sqrt{2}$.

$\langle C^1_{+1} \rangle = \langle C^1_{-1} \rangle = \langle C^1_0 \rangle/\sqrt{2}$.

$\langle C^2_0 \rangle = 0.$  $\langle C^2_{+1} \rangle = -\langle C^2_{-1} \rangle = (1/4)[\langle T^2_{+1} - T^2_{-1} \rangle + \langle T^2_{+2} - T^2_{-2} \rangle]$.

$\langle C^2_{+2} \rangle = -\langle C^2_{-2} \rangle = \langle C^2_{+1} \rangle$.

f = 2.  $\langle C^1_0 \rangle = \langle C^1_{+1} \rangle = \langle C^1_{-1} \rangle = 0$.

$\langle C^2_0 \rangle = (1/4)\sqrt{(3/2)} \langle T^2_{+1} + T^2_{-1} \rangle + (1/8)\sqrt{(3/2)} \langle T^2_{+2} + T^2_{-2} \rangle + (3/8) \langle T^2_0 \rangle$.

$\langle C^2_{+1} \rangle = \langle C^2_{-1} \rangle = \sqrt{(2/3)} \langle C^2_0 \rangle$.  $\langle C^2_{+2} \rangle = \langle C^2_{-2} \rangle = (1/2) \langle C^2_{+1} \rangle$.

**Table II**. Inferred values of multipole contributions to expressions (7) and (9). Values of $A_{K,Q}$ and $B_{K,Q}$ for 10 K.

|  | $A_{1,0}$ | $B_{1,1}$ | $B_{2,1}$ | $A_{2,0}$ | $A_{2,2}$ |
|---|---|---|---|---|---|
| $(0, \tau, 0)$ | 1.0 | $-i0.45$ |  |  |  |
| $(0, 2\tau, 0)$ |  |  | $i0.71$ | 1.0 | 0.06 |